\documentclass[12pt]{iopart}

\usepackage{amssymb}
\usepackage{mathenv}
\usepackage{graphicx}

\newcommand{\F}{\mathbf{F}}

\begin{document}

\title{The spacetime in the neighborhood of a general isolated black
  hole} 

\author{Badri Krishnan$^{1,2}$} 

\address{$^1$Max Planck Institute for Gravitational Physics, Albert
  Einstein Institute, Am M\"uhlenberg 1, D-14476 Potsdam, Germany }

\address{$^2$Max Planck Institute for Gravitational Physics, Albert
  Einstein Institute, Callinstrasse 38, D-30167 Hannover, Germany}

\ead{badri.krishnan@aei.mpg.de}

\begin{abstract}
  We construct the spacetime in the vicinity of a general isolated,
  rotating, charged black hole.  The black hole is modeled as a
  weakly isolated horizon, and we use the characteristic initial value
  formulation of the Einstein equations with the horizon as an inner
  boundary.  The spacetime metric and other geometric fields are
  expanded in a power series in a radial coordinate away from the
  horizon by solving the characteristic field equations in the
  Newman-Penrose formalism.  This is the first in a series of papers
  which investigate the near horizon geometry and its physical
  applications using the isolated horizon framework.
\end{abstract}

\section{Introduction}
\label{sec:intro}

The intrinsic geometry of a classical black hole horizon in
equilibrium is well understood.  The most general treatment is
provided by the framework of isolated horizons which allows for the
possibility of a black hole with arbitrary (but time independent)
intrinsic geometry in an otherwise dynamical spacetime
\cite{Ashtekar:1998sp,Ashtekar:1999yj,Ashtekar:2000hw,Ashtekar:2000sz,Ashtekar:2001is,Ashtekar:2001jb,Adamo:2009cd}.
Isolated horizons have been applied in various physical
circumstances. Some illustrative examples are black hole
thermodynamics,
black hole entropy calculations in quantum gravity, 
numerical relativity, 
and hairy black holes.  
More general situations when the black grows due to in-falling matter
and/or radiation have also been studied; see
e.g. \cite{Hayward:1993wb,Ashtekar:2002ag}.  See
\cite{Ashtekar:2004cn,Booth:2005qc,Gourgoulhon:2005ng,Hayward:2008ti} for reviews and
references.

In order to use isolated horizons in other astrophysical scenarios, it
is important to calculate the metric in some neighborhood of the black
hole.  Thus it is interesting to solve the Einstein equations in the
neighborhood of a black hole horizon.  Such calculations have been
carried out in the astropysically interesting context of a
Schwarzschild or Kerr black hole which is tidally deformed due to its
environment
\cite{Comeau:2009bz,Poisson:2009qj,Poisson:2005pi,Yunes:2005ve,Damour:2009va}.
Our approach here is closest to the work of Poisson and collaborators
\cite{Poisson:2009qj,Poisson:2005pi} who study the tidal deformation
of a non-rotating black hole using coordinates based on the past light
cones originating from the horizon. These coordinates were originally
described in \cite{Ashtekar:2000sz} and are similar to the Bondi
coordinates near null infinity \cite{Bondi:1962px}.  The present paper
goes towards generalizing \cite{Poisson:2009qj,Poisson:2005pi} to
include rotation, electric charge, and higher multipoles.

We use the characteristic initial value formulation of Einstein's
equations where free data is specified on a set of intersecting null
hyper-surfaces
\cite{Friedrich:1981at,Rendall:1990,Friedrich:2000qv,Stewart:1991}.
Consider $N$ dependent variables $\psi_I (I=1,\ldots,N)$ on a
spacetime manifold with coodinates $x^a$. We shall be concerned with
hyperbolic first-order quasilinear equations of the form
\begin{equation}
  \sum_{J=1}^N A^a_{IJ}(x,\psi)\partial_a\psi_J + F_I(x,\psi) = 0\,.
\end{equation}
In the standard Cauchy problem, one specifies the $\psi_I$ at some
initial time.  A solution is then guaranteed to be unique and to exist
at least locally in time.  The characteristic formulation considers
a pair of null surfaces $\mathcal{N}_0$ and $\mathcal{N}_1$ whose
intersection is a co-dimension-2 spacelike surface $S$.  It turns out
to be possible to specify appropriate data on the null surfaces and on
$S$ such that the above system of equations is well posed and has a
unique solution, at least locally near $S$.

\begin{figure}
  \begin{center}
    \includegraphics[angle=-0,width=0.8\textwidth]{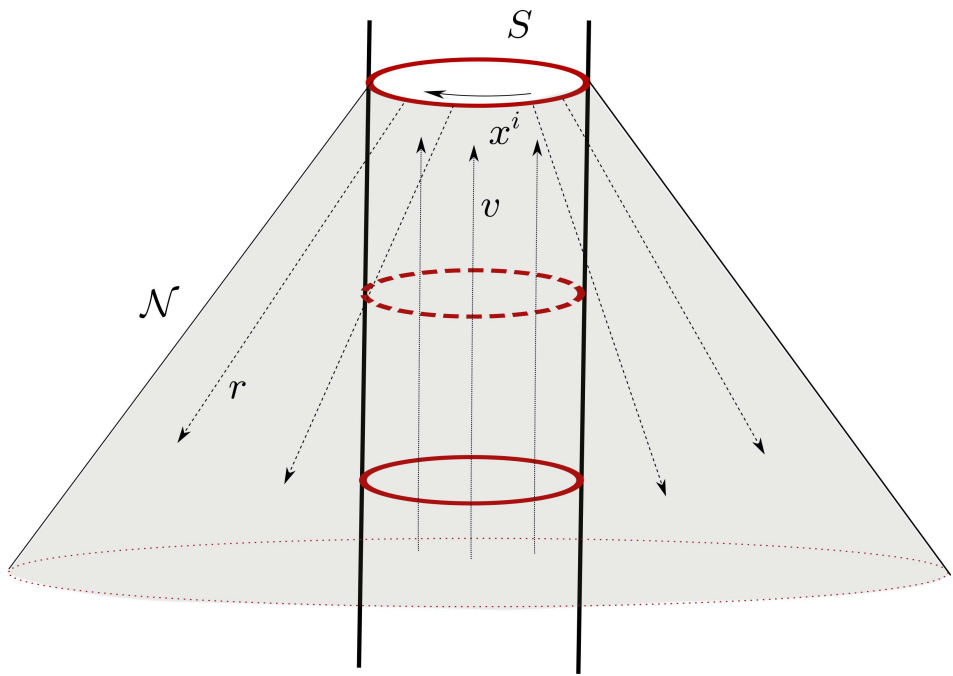}
    \caption{The near horizon coordinate system. The null generators
      of the horizon are parameterized by $v$, surfaces of constant
      $v$ are spheres.  On any such sphere choose coordinates $x^i$.
      The past directed outgoing light cone originating from $S$ is
      $\mathcal{N}$ and $r$ an affine parameter along the null
      generators of $\mathcal{N}$. }
    \label{fig:fig1}
  \end{center}
\end{figure}

In our case, the appropriate free data is specified on the horizon and
on an outgoing past light cone originating from a cross section of the
horizon.  Such a construction in the context of isolated horizons was
first studied by Lewandowski \cite{Lewandowski:1999zs} who
characterized the general solution of Einstein equations admitting an
isolated horizon.  The general scenario is sketched in
Figure~\ref{fig:fig1}.  We consider a portion of the horizon $\Delta$
which is isolated, in the sense that no matter and/or radiation is
falling into this portion of the horizon.  For a cross-section $S$,
the past-outgoing light cone is denoted by $\mathcal{N}$. The null
generators of $\Delta$ and $\mathcal{N}$ are parameterized by $v$ and
$r$ respectively; $x^i$ are coordinates on $S$.  This leads to a
coordinate system $(v,r,x^i)$ which is valid till the null geodesics
on $\mathcal{N}$ start to cross.  The field equations are solved in a
power series in $r$ away from the horizon.  This construction will be
spelled out more precisely in the course of this paper.

Unlike in \cite{Poisson:2009qj,Poisson:2005pi} we use the
Newman-Penrose formalism \cite{Newman:1961qr} which, as we shall see,
is well suited to this problem because of the central role played by
null surfaces.  Furthermore, including rotation and electric charge do
not make the equations much more complicated in the Newman-Penrose
framework.  The results of this paper cannot yet be directly compared
with \cite{Poisson:2009qj,Poisson:2005pi}.  This would require us to
start with, say a Kerr black hole, perturb it by considering a
non-vanishing background curvature, and to expand the perturbation in
powers of $r$.  Perturbation theory turns out to be straightforward in
this general set-up, at least in principle. This will be studied in a
forthcoming paper.  We make no attempt to study the issue of global
existence of solutions, and our solutions are valid only in a
neighborhood of $\Delta$.  There is no guarantee, for example, that
our solutions could be extended out to an asymptotically flat
region. Nevertheless, we expect that our construction does include
most solutions of possible astrophysical relevance.  In fact, it can
be shown numerically \cite{Diener} that for a Kerr black hole, this
coordinate system extends all way out to past null infinity.  It is
thus reasonable to expect that for perturbations of Kerr, the
coordinate system extends sufficiently far away from the horizon.  

The plan for this paper is as follows.  Section~\ref{sec:ih} reviews
the Newman-Penrose formalism and summarizes the definitions and some
basic properties of non-expanding and weakly isolated horizons.  This
will form the basis of the inner boundary conditions that will be
imposed later.  Section \ref{sec:coord} sets up the near horizon
coordinate system and gauge conditions.  We start by ignoring matter
fields which will be included later in Section \ref{sec:maxwell}.
Section \ref{sec:np} summarizes the Einstein field equations and the
Bianchi identities for vacuum spacetimes in the Newman-Penrose
formalism, and Sec.~\ref{sec:radial} explicitly solves the field
equations in powers of the radial coordinate $r$. Section
\ref{sec:maxwell} considers Einstein-Maxwell theory by incorporating
the source free Maxwell equations.  This section points out the
specific aspects of the previous calculations which need to be
modified due to the presence of an electromagnetic field.  Finally
Sec.~\ref{sec:conc} concludes with a summary and suggestions for
further investigations.

For simplicity, all manifolds and geometric fields shall be assumed to
be smooth.  We work in units where $G=c=1$.  We use the abstract index
notation where lower case Latin letters $a,b,\ldots$ are 4-dimensional
spacetime indices, and $i,j,\ldots$ denote the 2-dimensional angular
directions.  We take the spacetime metric $g_{ab}$ to have a signature
$(-+++)$.  The Riemann tensor $R_{abcd}$ is defined by
$(\nabla_a\nabla_b-\nabla_b\nabla_a)X^c = -{R_{abd}}^cX^d$ where
$\nabla_a$ is the derivative operator compatible with $g_{ab}$ and
$X^a$ is an arbitrary smooth vector field.

\section{Basic notions}
\label{sec:ih}

In this section, for completeness and to set up notation, we briefly
review the Newman-Penrose formalism and the basic definitions and
properties of non-expanding and weakly isolated horizons.

\subsection{The Newman-Penrose formalism} 
\label{subsec:np}

The Newman-Penrose formalism \cite{Newman:1961qr,Newman:1962} is a
tetrad formalism where the tetrad elements are null vectors, which
makes it especially well suited for studying null surfaces.  See
\cite{Penrose:1985jw,Chandrasekhar:1985kt,Stewart:1991} for
pedagogical treatments (note that these references take the spacetime
metric to have a signature of $(+---)$ which is different from ours).
Start with a null tetrad $(\ell,n,m,\bar{m})$ wher $\ell$ and $n$ are
real null vectors, and $m$ is a complex null vector and $\bar{m}$ its
complex conjugate.  The tetrad is such that $\ell\cdot n = -1$,
$m\cdot \bar{m} = 1$, with all other inner products vanishing.  The
spacetime metric is thus given by
\begin{equation}
  \label{eq:51}
  g_{ab} = -\ell_an_b - n_a\ell_b + m_a\bar{m}_b + \bar{m}_am_b\,.
\end{equation}
Directional derivatives along the basis vectors are denoted as
\begin{equation}
  \label{eq:54}
  D := \ell^a\nabla_a\,,\quad \Delta := n^a\nabla_a\,,\quad \delta :=
  m^a\nabla_a\,,\quad \bar{\delta} := \bar{m}^a\nabla_a\,.
\end{equation}
(The symbol $\Delta$ is used for both the directional derivative along
$n^a$ and for the isolated horizon).  The components of the connection
are encoded in 12 complex scalars, the spin coefficients, defined via
the directional derivatives of the tetrad vectors:
\begin{EqSystem*}[eq:spincoeffs]
  D\ell& = (\epsilon + \bar{\epsilon})\ell - \bar{\kappa}m -
  \kappa\bar{m} \,,\\
  Dn& = -(\epsilon + \bar{\epsilon})n + \pi m + \bar{\pi}m \,,\\
  Dm& = \bar{\pi}\ell - \kappa n + (\epsilon - \bar{\epsilon})m \,,\\
  \Delta \ell& = (\gamma + \bar{\gamma})\ell - \bar{\tau}m -
  \tau\bar{m}\,,\label{eq:Deltal}\\
  \Delta n & = -(\gamma + \bar{\gamma})n + \nu m + \bar{\nu}\bar{m}\,,\label{eq:Deltan}\\
  \Delta m & = \bar{\nu}\ell - \tau n + (\gamma-\bar{\gamma})m\,,\label{eq:Deltam}\\
  \delta\ell & = (\bar{\alpha} + \beta)\ell -\bar{\rho}m -
  \sigma\bar{m}\,,\\
  \delta n & = -(\bar{\alpha} + \beta)n + \mu m +
  \bar{\lambda}\bar{m}\,,\\
  \delta m & = \bar{\lambda}\ell - \sigma n + (\beta-\bar{\alpha})m\,,\label{eq:deltam}\\
  \bar{\delta}m & = \bar{\mu}\ell - \rho n + (\alpha-\bar{\beta})m\,. \label{eq:deltabarm}
\end{EqSystem*}
Many of the spin coefficients have a transparent geometric
interpretation.  Some important ones for us are: the real parts of
$\rho$ and $\mu$ are the expansion of $\ell$ and $n$ respectively;
$\sigma$ and $\lambda$ are the shears of $\ell$ and $n$ respectively;
the vanishing of $\kappa$ and $\nu$ implies that $\ell$ and $n$ are
respectively geodesic; $\epsilon + \bar{\epsilon}$ and $\gamma +
\bar{\gamma}$ are respectively the accelerations of $\ell$ and $n$,
$\alpha-\bar{\beta}$ yields the connection in the $m$-$\bar{m}$ plane
and thus the curvature of the manifold spanned by $m$-$\bar{m}$.

Since the null tetrad is typically not a coordinate basis, the above definitions
of the spin coefficients lead to non-trivial commutation relations:
\begin{EqSystem*}[eq:commutation]
  (\Delta D - D\Delta)f &= (\epsilon+\bar{\epsilon})\Delta f + (\gamma
  + \bar\gamma)Df - (\bar\tau + \pi)\delta f -(\tau + \bar{\pi})\bar{\delta}f\,,\\
  (\delta D-D\delta)f &= (\bar{\alpha}+\beta-\bar\pi)Df + \kappa\Delta
  f - (\bar{\rho}+\epsilon-\bar{\epsilon})\delta f
  -\sigma\bar{\delta}f\,,\\
  (\delta\Delta-\Delta\delta)f &= -\bar\nu Df +
  (\tau - \bar{\alpha}-\beta)\Delta f + (\mu-\gamma+\bar\gamma)\delta f + \bar{\lambda}\bar{\delta}f\,,\\
  (\bar{\delta}\delta-\delta\bar{\delta})f &= (\bar{\mu}-\mu)Df +
  (\bar{\rho}-\rho)\Delta f + (\alpha-\bar{\beta})\delta f -
  (\bar{\alpha}-\beta)\bar{\delta}f\,.
\end{EqSystem*}
The Weyl tensor $C_{abcd}$ breaks down into 5 complex scalars
\begin{EqSystem*}
  \Psi_0 = C_{abcd}\ell^am^b\ell^cm^d\,,\quad
  \Psi_1 = C_{abcd}\ell^am^b\ell^cn^d\,,\quad
  \Psi_2 = C_{abcd}\ell^am^b\bar{m}^cn^d\,,\\
  \Psi_3 = C_{abcd}\ell^an^b\bar{m}^cn^d\,,\quad
  \Psi_4 = C_{abcd}\bar{m}^an^b\bar{m}^cn^d\,.
\end{EqSystem*}
Similarly, the Ricci tensor is decomposed into 4 real and 3 complex
scalars $\Phi_{ij}$: 
\begin{EqSystem*}[eq:riccicomp]
  \Phi_{00} = \frac{1}{2}R_{ab}\ell^a\ell^b \,,\quad
  \Phi_{11} = \frac{1}{4}R_{ab}(\ell^a n^b + m^a \bar{m}^b) \,,\quad
  \Phi_{22} = \frac{1}{2}R_{ab}n^a n^b \,,\quad
  \Lambda = \frac{R}{24}\,,\\
  \Phi_{01} = \frac{1}{2}R_{ab}\ell^a m^b \,,\quad
  \Phi_{02} = \frac{1}{2}R_{ab}m^a m^b \,,\quad
  \Phi_{12}  = \frac{1}{2}R_{ab}m^a n^b \,,\quad
  \bar{\Phi}_{ij} = \Phi_{ji}\,.
\end{EqSystem*}
The 6 components of the Maxwell 2-form $\F_{ab}$ are written in terms
of 3 complex scalars
\begin{equation}
  \phi_0 = -\F_{ab}\ell^a m^b\,,\quad
  \phi_1 = \frac{1}{2}\F_{ab}(n^a\ell^b + m^a\bar{m}^b)\,,\quad 
  \phi_2 = \F_{ab}n^a\bar{m}^b \,.
\end{equation}
The source free Maxwell equations $d\F = 0$ and $d\star\F = 0$ are
written as 4 complex scalar equations
\begin{EqSystem*}[eq:maxwelleqs]
  D\phi_1 - \bar{\delta}\phi_0 = (\pi-2\alpha)\phi_0 + 2\rho\phi_1 -
  \kappa\phi_2\,,\\
  D\phi_2 - \bar{\delta}\phi_1 = -\lambda\phi_0 + 2\pi\phi_1 +
  (\rho-2\epsilon)\phi_2 \,,\\
  \Delta\phi_0 - \delta\phi_1 = (2\gamma-\mu)\phi_0 - 2\tau\phi_1 +
  \sigma\phi_2 \,,\\
  \Delta\phi_1 - \delta\phi_2 = \nu\phi_0 - 2\mu\phi_1 +
  (2\beta-\tau)\phi_2\,. 
\end{EqSystem*}
The stress energy tensor for the Maxwell field is given by 
\begin{equation}
  T_{ab} = \frac{1}{4\pi}\left(\F_{ac}{\F_{b}}^c - \frac{1}{4}g_{ab}\F_{cd}\F^{cd} \right)\,, 
\end{equation}
which is seen to be trace-free.  Using the Einstein equations, we see
that for Einstein-Maxwell theory, the Ricci tensor components defined
in Eqs.~(\ref{eq:riccicomp}) have a simple expression in terms of the
$\phi_i$:
\begin{equation}
  \label{eq:5}
  \Phi_{ij} = 2\phi_i\bar{\phi}_j \,\, (i,j=0,1,2)\,,\quad \Lambda = 0\,.
\end{equation}
The relation between the spin coefficients and the curvature
components lead to the so called Newman-Penrose field equations which
are a set of 16 complex first order differential equations.  The
Bianchi identities are written explicitly as 8 complex and 3 real
equations.  See
\cite{Penrose:1985jw,Chandrasekhar:1985kt,Stewart:1991} for the full
set of field equations and Bianchi identities.  We shall later write
these equations after imposing gauge and coordinate conditions.

It will also be useful to use the notion of spin weights and the
$\eth$ operator for derivatives in the $m$-$\bar{m}$ plane (which will
be angular derivatives in our case).  A tensor $X$ projected on the
$m$-$\bar{m}$ plane is said to have spin weight $s$ if under a spin
rotation $m\rightarrow e^{i\psi}m$, it transforms as $X\rightarrow
e^{i s\psi}X$.  Thus, $m^a$ itself has spin weight $+1$ while
$\bar{m}^a$ has weight $-1$.  For a scalar $X= m^{a_1}\cdots
m^{a_p}\bar{m}^{b_1}\cdots \bar{m}^{b_q}X_{{a_1}\cdots {b_q}}$,
i.e. it has $p$ contractions with $m$ and $q$ with $\bar{m}$, then $X$
has spin weight $s=p-q$.  For example, the Weyl tensor component
$\Psi_k$ has spin weight $2-k$.  Similarly, the Maxwell field
component $\phi_k$ has weight $1-k$.

The $\eth$ and $\bar{\eth}$ operators are defined as
\begin{eqnarray}
  \label{eq:19}
  \eth X = m^{a_1}\cdots m^{a_p}\bar{m}^{b_1}\cdots
  \bar{m}^{b_q}\delta X_{{a_1}\cdots {b_q}}\,,\\
  \bar{\eth} X = m^{a_1}\cdots m^{a_p}\bar{m}^{b_1}\cdots
  \bar{m}^{b_q}\bar{\delta} X_{{a_1}\cdots {b_q}}\,.
\end{eqnarray}
From Eqs.~(\ref{eq:deltam}) and (\ref{eq:deltabarm}), after projecting
on to the $m$-$\bar{m}$ plane, we get
\begin{equation}
  \delta m^a = (\beta-\bar{\alpha})m^a\,,\qquad \bar{\delta}{m}^a =
  (\alpha-\bar{\beta})m^a\,. 
\end{equation}
A short calculation shows that
\begin{equation}
  \label{eq:55}
  \eth X = \delta X + s(\bar{\alpha}-\beta)X\,,\qquad \bar{\eth}X =
  \bar{\delta}X - s(\alpha-\bar{\beta})X\,.
\end{equation}
It is clear that $\eth$ and $\bar{\eth}$ act as spin raising and
lowering operators.  See \cite{Goldberg:1966uu} for further properties
of the $\eth$ operator and its connection to representations of the
rotation group.  

The transformations of the null tetrad which preserve the metric are
\begin{description}
\item[(i)] the boosts:
\begin{equation}
  \label{eq:32}
  l\rightarrow Al\,,\quad n\rightarrow A^{-1}n\,,\quad m\rightarrow m\,,
\end{equation}

\item[(ii)] the spin transformations in the $m-\bar{m}$ plane: 
\begin{equation}
  \label{eq:31}
  m\rightarrow e^{i\psi}m\,,\quad \ell\rightarrow \ell\,,\quad
  n\rightarrow n\,,
\end{equation}
\item[(iii)] the null rotations around $\ell$:
\begin{equation}
  \label{eq:53}
  \ell\rightarrow \ell\,,\quad m\rightarrow m + a\ell\,,\quad n
  \rightarrow n + \bar{a}m + a\bar{m} + |a|^2\ell\,,
\end{equation}
\item[(iv)] the null rotations around $n$ (obtained by interchanging
$\ell$ and $n$ in Eq.~(\ref{eq:53})).  
\end{description}
Again, we refer to
\cite{Penrose:1985jw,Chandrasekhar:1985kt,Stewart:1991} for a more
complete discussion.

\subsection{Non-expanding and weakly isolated horizons}

A black hole in equilibrium with its surroundings is modeled
quasi-locally as an isolated horizon.  The basic geometrical object is
a smooth 3-dimensional null surface $\Delta$ (which shall be the black
hole horizon) in a Lorentzian spacetime $(\mathcal{M}$, $g_{ab})$. If
$\ell$ is any null normal of $\Delta$, then it must be geodesic so
that $\ell^a\nabla_a\ell^b = \tilde{\kappa}_{(\ell)}\ell^b$; the
acceleration $\tilde{\kappa}_{(\ell)}$ is the \emph{surface gravity}
associated with $\ell^a$.  We shall consider only non-extremal
horizons here, i.e. we shall always have non-vanishing
$\tilde{\kappa}$. The spacetime metric $g_{ab}$ induces a degenerate
metric on $\Delta$ which we denote $q_{ab}$ which has signature
$(0++)$.  Its (non-unique) inverse will be denoted $q^{ab}$.  There is
a volume element ${}^2\epsilon_{ab}$ on $\Delta$, satisfying
$\ell^a\,{}^2\epsilon_{ab} = 0$, which measures the area of spacelike
cross-sections of $\Delta$.

A smooth 3-dimensional null surface $\Delta$ is said to be a
\emph{non-expanding horizon} if:
\begin{itemize}
\item $\Delta$ has topology $S^2\times\mathbb{R}$.
\item The expansion $\Theta_{(\ell)} := q^{ab}\nabla_a\ell_b$ of any
  null normal $\ell^a$ of $\Delta$ vanishes.
\item The Einstein field equations hold at $\Delta$, and the matter
  stress-energy tensor $T_{ab}$ is such that for any future directed
  null-normal $\ell^a$, $-T^a_b\ell^b$ is future causal.  
\end{itemize}
We shall consider only null tetrads adapted to $\Delta$ such that, at
the horizon, $\ell^a$ coincides with a null-normal to $\Delta$.  We
shall also consider a foliation of the horizon by spacelike spheres
$S_v$ with $v$ a coordinate on the horizon which is also an affine
parameter along $\ell$: $\mathcal{L}_\ell v = 1$.  Null rotations
about $\ell^a$ correspond to changing the foliation.

This deceptively simple definition of a non-expanding horizon leads to
a number of important results which we state here without proof
(though some of these will be rederived later):
\begin{enumerate}
\item The Weyl tensor components $\Psi_0$ and $\Psi_1$ vanish on the
  horizon.  This implies that $\Psi_2$ is an invariant on $\Delta$ as
  long as the null-tetrad is adapted to the horizon; it is
  automatically invariant under boosts and spin rotations (it has spin
  weight $0$), and it is invariant under null rotations around $\ell$
  because $\Psi_0$ and $\Psi_1$ vanish.  Similarly, the Maxwell field
  component $\phi_0$ vanishes on the horizon, and $\phi_1$ is
  invariant on $\Delta$.  Both $\Psi_2$ and $\phi_1$ are also time
  independent on the horizon.    
\item For a general null sub-manifold, there is no unique derivative
  operator compatible with the metric, and the pull back of the
  spacetime derivative operator $\nabla_a$ does not necessarily induce
  a connection on the hypersurface.  For non-expanding horizons
  however, the spacetime connection does induce a unique derivative
  operator compatible with $q_{ab}$. Furthermore, there exists a
  1-form $\omega_a$ such that, for any vector field $X^a$ tangent to
  $\Delta$,
  \begin{equation}
    \label{eq:14}
    X^a\nabla_a\ell^b = X^a\omega_a\ell^b\,.
  \end{equation}
  The 1-form $\omega_a$ plays a fundamental role in what follows.  The
  pullback of $\omega_a$ to the cross-sections $S$ will be denoted
  $\tilde{\omega}_a$.
\item The surface gravity of $\ell$ is
  \begin{equation}
    \label{eq:15}
    \tilde{\kappa}_{(\ell)} = \ell^a\omega_a\,.
  \end{equation}
  The curl and divergence of $\omega$ carry important physical
  information.  The curl is related to the imaginary part of the Weyl
  tensor on the horizon
  \begin{equation}
    \label{eq:16}
    d\omega = \textrm{Im}\left[\Psi_2\right] {}^{2}\epsilon\,.
  \end{equation}
  and its divergence specifies the foliation of $\Delta$ by spheres
  \cite{Ashtekar:2001jb}.

\item The horizon angular momentum is well defined in the case when
  there is an axial symmetry $\varphi^a$ on $\Delta$ which preserves
  $q_{ab}$, $\omega_a$ and the electromagnetic field on the horizon
  \cite{Ashtekar:2001is}.  The angular momentum is given by
  \begin{equation}
    J  = -\frac{1}{4\pi}\oint_S f\mathrm{Im}[\Psi_2]\,{}^2\epsilon +
    \frac{1}{2\pi}\oint_S g\mathrm{Im}[\phi_1]\,{}^2\epsilon\,, 
  \end{equation}
  where $f$ and $g$ are respectively defined via
  $\varphi^a{}^2\epsilon_{ab} = \partial_bf$ and $\varphi^a\star
  \mathbf{F}_{ab} = \partial_b g$.  Similarly, the electric and
  magnetic charges of the horizon are defined respectively as
  \begin{equation}
    \label{eq:emcharges}
    Q = \frac{1}{2\pi}\oint_S \mathrm{Re}\left[\phi_1\right]
    \,{}^2\epsilon\,,\quad   P = \frac{1}{2\pi}\oint_S \mathrm{Im}\left[\phi_1\right]
    \,{}^2\epsilon\,.
  \end{equation}
  Hamiltonian methods provide a suitable notion of horizon mass
  \cite{Ashtekar:2001is}:
  \begin{equation}
    M = \frac{1}{2R} \sqrt{(R^2+Q^2)^2 + 4J^2}\,,
  \end{equation}
  where $R$ is the area radius of the horizon so that if $A$ is the
  area of the horizon cross-sections then $R := \sqrt{A/4\pi}$. In the case
  when the horizon is not exactly symmetric, one could attempt to find
  an approximate symmetry to replace $\varphi^a$ in the above
  equations \cite{Dreyer:2002mx,Beetle:2008yt,Cook:2007wr}.  It is
  also possible to define source multipole moments for an
  axisymmetric charged isolated horizon \cite{Ashtekar:2004gp}.  

\end{enumerate}
We need to strengthen the conditions of a non-expanding horizon for
various physical situations.  The minimum extra condition required for
black hole thermodynamics and to have a well defined action principle
with $\Delta$ as an inner boundary of a portion of spacetime, is
formulated as a weakly isolated horizon \cite{Ashtekar:2000hw}.  This
is to choose an equivalence class of null normals $[\ell]$, each
related to the other by a constant re-scaling $\ell^\prime = c\ell$,
such that
\begin{equation}
  \label{eq:39}
  \mathcal{L}_\ell\omega_a = 0\,.
\end{equation}
This can be shown to be equivalent to the zeroth law,
i.e. $\tilde{\kappa}_{(\ell)} = \ell^a\omega_a$ is constant on the
horizon.  Note that under a re-scaling $\ell^a \rightarrow f\ell^a$,
$\omega_a$ transforms as $\omega_a \rightarrow \omega_a + D_a\ln f$ so
that it is invariant under constant rescalings.  This condition is
sufficient to ensure a well defined action principle and to lead to a
sensible notion of horizon mass and spin.  Any non-expanding horizon
can be made into a weakly isolated horizon by suitably scaling the
null generators.  Thus, the restriction to weakly isolated horizons is
not a genuine physical restriction.

\section{The near-horizon coordinate system and null-tetrad}
\label{sec:coord}

Let us now assume that the vacuum Einstein equations hold in a
neighborhood of the horizon $\Delta$.  We will consider
electromagnetic fields later in Sec.~\ref{sec:maxwell}.  Following
\cite{Ashtekar:2000sz} we introduce a coordinate system and null
tetrad in the vicinity of $\Delta$ analogous to the Bondi coordinates
near null infinity.  See Fig.~\ref{fig:fig1}.  Choose a particular
null normal $\ell^a$ on $\Delta$. Let $v$ be the affine parameter
along $\ell^a$ so that $\ell^a\nabla_a v = 1$.  Let $S_v$ denote the
spheres of constant $v$. Introduce coordinates $x^i$ ($i=2,3$) on any
one $S_v$ (call this sphere $S_0$) and require them to be constant
along $\ell^a$: $\ell^a\nabla_a x^i = 0$; this leads to a coordinate
system $(v,x^i)$ on $\Delta$.  Let $n^a$ be a future directed inward
pointing null vector orthogonal to the $S_v$ and normalized such that
$\ell\cdot n = -1$. Extend $n^a$ off $\Delta$ geodesically, with $r$
being an affine parameter along $-n^a$; set $r=0$ at $\Delta$.  This
yields a family of null surfaces $\mathcal{N}_v$ parameterized by $v$
and orthogonal to the spheres $S_v$.  Set $(v,x^i)$ to be constant
along the integral curves of $n^a$ to obtain a coordinate system
$(v,r,x^i)$ in a neighborhood of $\Delta$.  Choose a complex null
vector $m^a$ tangent to $S_0$.  Lie drag $m^a$ along $\ell^a$:
\begin{equation}
  \label{eq:17}
  \mathcal{L}_\ell m^a = 0 \,\quad \textrm{on}\,\Delta\,.
\end{equation}
We thus obtain a null tetrad $(\ell,n,m,\bar{m})$ on
$\Delta$. Finally, parallel transport $\ell$ and $m$ along $-n^a$ to
obtain a null tetrad in the neighborhood of $\Delta$.  This
construction is fixed up to the choice of the $x^i$ and $m^a$ on an
initial cross-section $S_0$.  We are allowed to perform an arbitrary
spin transformation $m\rightarrow e^{i\psi}m$ on $S_0$.

It is easy to write the elements null tetrad in these $(v,r,x^i)$
coordinates.  Let us start with $n_a$ and $n^a$.  We have the family of
null surfaces $\mathcal{N}_v$ parameterized by $v$; $n_a$ is
normal to the $\mathcal{N}_v$, and $r$ is an affine parameter along
$-n^a$.  This implies that we can choose
\begin{equation}
  \label{eq:2}
  n_a = -\partial_a v \qquad \textrm{and} \qquad n^a\nabla_a
  :=\Delta = -\frac{\partial}{\partial r}\,.
\end{equation}
To satisfy the inner-product relations, the other basis vectors must
be of the form:
\begin{equation}
  \label{eq:3}
  \ell^a\nabla_a := D = \frac{\partial}{\partial v} + U\frac{\partial}{\partial r}
  + X^i\frac{\partial}{\partial x^i}\,,\quad m^a\nabla_a:= \delta =
  \Omega\frac{\partial}{\partial r} + \xi^i\frac{\partial}{\partial x^i}\,.
\end{equation}
The frame functions $U,X^i$ are real while $\Omega,\xi^i$ are complex.
We want $\ell^a$ to be a null normal of $\Delta$ so that the null
tetrad is adapted to the horizon.  Since $\partial_v$ is tangent to
the null generators of $\Delta$, this clearly requires that $U,X^i$
must vanish on the horizon.  Similarly, we want $m^a$ to be tangent to
the spheres $S_v$ at the horizon, so $\Omega$ should also vanish on
$\Delta$.  Thus, $U,X^i,\Omega$ are all $\mathcal{O}(r)$ functions.

We expand the spin coefficients, Weyl tensor components and the
directional derivatives in a power series in $r$ away from the horizon
\begin{equation}
  X = X^{(0)} + rX^{(1)} + \frac{1}{2}r^2 X^{(2)} \ldots\,.
\end{equation}
Thus, for example we will have
\begin{equation}
  \label{eq:35}
  \Psi_k = \Psi_k^{(0)} + r\Psi_k^{(1)} + \frac{1}{2}r^2\Psi_k^{(2)} + \ldots\,. 
\end{equation}
The same notation will be used for the frame fields $U$ and $\Omega$.
However, for the frame fields $X^i$ and $\xi^i$, we will write
\begin{equation}
  \xi^i = \xi^i_{(0)} + r\xi^i_{(1)} + \frac{1}{2}r^2\xi^i_{(2)} + \ldots\,.
\end{equation}
To avoid clutter we shall not be completely consistent with this
notation.  Thus, we shall not use any index for the spin weighted
angular derivative $\eth$; it is to be understood that $\eth$ always
refers to $\eth^{(0)}$ in this paper. Similarly, where we don't expect
any confusion, we shall often drop the index on the directional
derivatives such as $D$ and $\delta$; in this case, unless mentioned
otherwise, the relevant order of the operator is the same as the order
of the operand.  For example, $\delta\Psi_2^{(0)}$ refers to
$\delta^{(0)}\Psi_2^{(0)}$.

The variables we need to solve for are the frame fields
$U,X^i,\Omega,\xi^i$, the 12 spin coefficients and the Weyl tensor
components $\Psi_k$.  The equations are the commutations relations
Eqs.~(\ref{eq:3}), the 16 field equations and 8 of the 11 Bianchi
identities.  Three of the Bianchi identities involve only the Ricci
tensor, thus they will need to be considered when matter fields are
present.  In Sec.~\ref{sec:maxwell} we will consider the Maxwell
equations and the 3 additional Bianchi identities as well. All these
are first order differential equations and each of these sets of
equations has possibly three kinds of equations: evolution equations
which involve derivatives along $v$, i.e. $D$, and do not contain any
radial derivatives $\Delta$, equations which contain only purely
angular derivatives $\delta$ and $\bar\delta$, and finally the radial
equations involving $\Delta$.  In order to integrate these equations,
we proceed as follows. We start with suitable data on some initial
cross section $S_0$ of the horizon and use the non-radial equations to
propagate them at all points of the horizon.  Starting with this
horizon data and the appropriate data on the past light cone
$\mathcal{N}_0$ containing $S_0$, the radial equations then yield the
first radial derivatives and, iteratively, all successive higher
derivatives as well.  At each step, we will need to expand the
non-radial equations in powers of $r$, and ensure that we have
consistency order-by-order in $r$.  We now carry out this procedure in
detail and spell out the free data.

We start with the conditions on the spin coefficients.  Since $n^a$ is
an affinely parameterized geodesic, and $\ell$ and $m$ are parallel
propagated along $n^a$, we have $\Delta n = \Delta \ell = \Delta m =
0$.  From Eqs.~(\ref{eq:Deltal}), (\ref{eq:Deltan}) and
(\ref{eq:Deltam}), this leads to
\begin{equation}
  \label{eq:13}
  \gamma = \tau = \nu = 0\,.
\end{equation}
Imposing Eq.~(\ref{eq:13}) in the commutation relations (\ref{eq:commutation})
leads to
\begin{EqSystem*}[eq:commutation1]
  (\Delta D - D\Delta)f &= (\epsilon+\bar{\epsilon})\Delta f - \pi\delta f -
  \bar{\pi}\bar{\delta}f\,,\\
  (\delta D-D\delta)f &= (\bar{\alpha}+\beta-\bar\pi)Df + \kappa\Delta f -
  (\bar{\rho}+\epsilon-\bar{\epsilon})\delta f
  -\sigma\bar{\delta}f\,,\\
  (\delta\Delta-\Delta\delta)f &=
  -(\bar{\alpha}+\beta)\Delta f + \mu\delta f + \bar{\lambda}\bar{\delta}f\,,\\
  (\bar{\delta}\delta-\delta\bar{\delta})f &= (\bar{\mu}-\mu)Df +
  (\bar{\rho}-\rho)\Delta f + (\alpha-\bar{\beta})\delta f -
  (\bar{\alpha}-\beta)\bar{\delta}f\,. 
\end{EqSystem*}
Substituting $f=v$ yields
\begin{equation}
  \label{eq:6}
  \pi = \alpha + \bar{\beta}\,,\quad \mu = \bar{\mu}\,.
\end{equation}
Eqs.~(\ref{eq:13}) and (\ref{eq:6}) are the basic conditions on the
spin coefficients which hold at all spacetime points where the
coordinate system is valid.  In addition, we have the boundary
condition at the horizon, namely that $\ell$ is geodesic, normal to a
smooth surface, and expansion free. This implies
\begin{equation}
  \label{eq:36}
  \rho^{(0)} = \kappa^{(0)} = 0\,.
\end{equation}
The subscript ${}^{(0)}$ indicates that these are the values at the horizon,
i.e. at $r=0$.  From the Raychaudhuri equation applied to the null
generators of the horizon, we get 
\begin{equation}
  D\rho^{(0)} - \bar{\delta}\kappa^{(0)} = (\rho^{(0)})^2 +
  |\sigma^{(0)}|^2 + (\epsilon^{(0)} + \bar{\epsilon}^{(0)})\rho^{(0)}
  - 2\alpha^{(0)}\kappa^{(0)}
\end{equation}
(this is also one of the Newman-Penrose field equations considered
later).  This leads to $\sigma^{(0)}=0$.  In addition, we can make a
spin transformation on $S_0$ and set
\begin{equation}
  \label{eq:37}
  \epsilon^{(0)} - \bar{\epsilon}^{(0)} = 0\,.
\end{equation}
From Eq.~(\ref{eq:commutation1:b}) and the above conditions on the spin
coefficients at the horizon, this also ensures that $\mathcal{L}_\ell
m^a = 0$ at the horizon.  Finally, we assume that we have scaled the
null normal appropriately (i.e. we have chosen the coordinate $v$ and
the foliation $S_v$) so that the horizon is weakly isolated:
$\mathcal{L}_\ell \omega_a = 0$.  In terms of the Newman-Penrose spin
coefficients, $\omega_a$ is written as
\begin{equation}
  \label{eq:40}
  \omega_a = -n_b\nabla_a\ell^b = (\epsilon+\bar\epsilon) n_a + \pi^{(0)} m_a +
  \bar{\pi}^{(0)}\bar{m}_a\,. 
\end{equation}
Thus, $\Delta$ will be a weakly isolated horizon if we choose
\begin{equation}
  \label{eq:38}
  \tilde{\kappa} := \epsilon^{(0)} + \bar{\epsilon}^{(0)} =
  \textrm{constant}\,,\quad D\pi^{(0)} = 0\,.
\end{equation}
Among the free functions that we need to specify, we can choose a
constant for the surface gravity (perhaps determined by the mass and
spin \cite{Ashtekar:2001is,Ashtekar:2000hw}), and a function
$\pi^{(0)}$ on the initial cross-section $S_0$.

\section{The reduced system of equations}
\label{sec:np}

With the conditions Eqs.~(\ref{eq:13}) and (\ref{eq:6}) on the spin
coefficients at hand, we now impose them in the commutator relations,
the field equations and the Bianchi identities.  The functions
$U,X^i,\Omega,\xi^i$ are determined by the commutation relations
(\ref{eq:commutation1}) by substituting, in turn, $r$ and $x^i$ for
$f$, and imposing Eqs.~(\ref{eq:13}) and (\ref{eq:6}) on the spin
coefficients.  First the radial derivatives:
\begin{EqSystem*}[eq:4]
  \Delta U &= -(\epsilon+\bar{\epsilon}) - \pi\Omega -
  \bar{\pi}\bar{\Omega}\,,\\ 
  \Delta X^i &= -\pi\xi^i - \bar{\pi}\bar{\xi}^i\,,\\
  \Delta\Omega &= -\bar{\pi} - \mu\Omega -\bar{\lambda}\bar{\Omega}
  \,,\\ 
  \Delta\xi^i &= -\mu\xi^i -\bar{\lambda}\bar{\xi}^i\,.
\end{EqSystem*}
The propagation equations along $v$ are
\begin{EqSystem*}[eq:frametime]
   D\Omega - \delta U&= \kappa + \rho\Omega + \sigma\bar{\Omega}\,,\\ 
   D\xi^i - \delta X^i &= \bar{\rho}\xi^i + \sigma\bar{\xi}^i\,.
\end{EqSystem*}
Let us now turn to the field equations.  After imposing
Eqs.~(\ref{eq:13}) and (\ref{eq:6}) on the spin coefficients and
ignoring matter terms for now, the field equations involving radial
derivatives are:
\begin{EqSystem*}[eq:7]
  \Delta\lambda = -2\lambda\mu -\Psi_4\,,\\
  \Delta\mu = -\mu^2 - |\lambda|^2\,,\\
  \Delta\rho = -\mu\rho -\sigma\lambda -\Psi_2\,,\\
  \Delta\sigma = -\mu\sigma -\bar{\lambda}\rho\,.\\
  \Delta\kappa = -\bar{\pi}\rho - \pi\sigma -\Psi_1\,,\\
  \Delta\epsilon = -\bar{\pi}\alpha - \pi\beta - \Psi_2\,,\\
  \Delta\pi = -\pi\mu-\bar{\pi}\lambda -\Psi_3\,,\\
  \Delta\beta = -\mu\beta -\alpha\bar{\lambda}\,,\\
  \Delta\alpha = -\beta\lambda -\mu\alpha-\Psi_3\,.
\end{EqSystem*}
The time evolution equations become:
\begin{EqSystem*}[eq:8]
  D\rho -\bar{\delta}\kappa = \rho^2 + |\sigma|^2 +
  (\epsilon+\bar{\epsilon})\rho -2\alpha\kappa\,,\label{eq:drho}\\
  D\sigma-\delta\kappa = (\rho+\bar{\rho} +2\epsilon)\sigma
  -2\beta\kappa + \Psi_0\,,\label{eq:dsigma}\\
  D\alpha-\bar{\delta}\epsilon = (\rho-\epsilon)\alpha +
  \beta\bar{\sigma} - \bar{\beta}\epsilon -\kappa\lambda +
  (\epsilon+\rho)\pi\,,\\
  D\beta -\delta\epsilon = (\alpha+\pi)\sigma +
  (\bar{\rho}-\epsilon)\beta -\mu\kappa
  -(\bar{\alpha}-\bar{\pi})\epsilon + \Psi_1\,,\\
  D\lambda -\bar{\delta}\pi = (\rho-2\epsilon)\lambda +
  \bar{\sigma}\mu + 2\alpha\pi\,,\\
  D\mu - \delta\pi = (\bar\rho - 2\epsilon)\mu + \sigma\lambda +
  2\beta\pi + \Psi_2\,.
\end{EqSystem*}
The angular field equations are:
\begin{EqSystem*}[eq:9]
  \delta\rho - \bar\delta\sigma = \bar{\pi}\rho -
  (3\alpha-\bar{\beta})\sigma -\Psi_1\,,\\
  \delta\alpha - \bar{\delta}\beta = \mu\rho -\lambda\sigma +
  |\alpha|^2 + |\beta|^2 -2\alpha\beta -\Psi_2\,,\\
  \delta\lambda -\bar{\delta}\mu = \pi\mu +
  (\bar{\alpha}-3\beta)\lambda -\Psi_3\,.
\end{EqSystem*}
Finally, we have the Bianchi identities which, in the NP formalism,
are written as a set of nine complex and two real equations; in the
absence of matter, only 8 complex equations survive. The radial
Bianchi identities reduce to:
\begin{EqSystem*}[eq:10]
  \Delta\Psi_0 - \delta\Psi_1 = -\mu\Psi_0 -2\beta\Psi_1 +
  3\sigma\Psi_2\,,\\
  \Delta\Psi_1 -\delta\Psi_2 = -2\mu\Psi_1 + 2\sigma\Psi_3\,,\\
  \Delta\Psi_2 -\delta\Psi_3 = -3\mu\Psi_2 + 2\beta\Psi_3 +
  \sigma\Psi_4\,,\\ 
  \Delta\Psi_3 - \delta\Psi_4 = -4\mu\Psi_3 + 4\beta\Psi_4\,.
\end{EqSystem*}
Note that there is no equation for the radial derivative of $\Psi_4$.
Among all the fields that we are solving for, this is in fact the only
one for which this happens.  This means that $\Psi_4$ (in this case,
its radial derivatives) is the free data that must be specified on the
null cone $\mathcal{N}_0$ originating from $S_0$.

Finally, we have the components of the Bianchi equations for evolution
of the Weyl tensor components:
\begin{EqSystem*}[eq:11]
  D\Psi_1 -\bar\delta\Psi_0 = (\pi-4\alpha)\Psi_0 +
  2(2\rho+\epsilon)\Psi_1 -3\kappa\Psi_2\,,\\
  D\Psi_2-\bar\delta\Psi_1 = -\lambda\Psi_0 -2(\pi-\alpha)\Psi_1 +
  3\rho\Psi_2 -2\kappa\Psi_3\,,\\  
  D\Psi_3 -\bar\delta\Psi_2 = -2\lambda\Psi_1 + 3\pi\Psi_2
  +2(\rho-\epsilon)\Psi_3 -\kappa\Psi_4 \,,\\
  D\Psi_4-\bar\delta\Psi_3 = -3\lambda\Psi_2 +2(\alpha+2\pi)\Psi_3 +
  (\rho-4\epsilon)\Psi_4 \,.
\end{EqSystem*}
Having written down the full set of differential equations, we are now
ready to solve them locally near the horizon order-by-order in $r$.

\section{The intrinsic horizon geometry}
\label{sec:horizon}

We first consider the intrinsic geometry of the horizon contained in
the time evolution equations at the horizon, i.e. the
$\mathcal{O}(r^0)$ part of Eqs.~(\ref{eq:8}).  This has been studied
elsewhere in great detail
\cite{Ashtekar:2001jb,Ashtekar:2000hw,Adamo:2009cd}, so we shall be
brief.  We already have that $\ell^a$ is geodesic so that
$\kappa^{(0)}=0$, and the expansion and twist of $\ell$ vanish, which
implies $\rho^{(0)} = 0$.  The Eqs.~(\ref{eq:8}) in turn lead to
\begin{EqSystem*}[eq:12]
  \sigma^{(0)}=0\,,\qquad \Psi_0^{(0)}=0\,,\\
  D\alpha^{(0)}-\bar{\delta}\epsilon^{(0)} = \epsilon^{(0)}\alpha^{(0)}
  - \bar{\beta}^{(0)}\epsilon^{(0)}  + \epsilon^{(0)}\pi^{(0)}\,,\\
  D\beta^{(0)} -\delta\epsilon^{(0)} = (\alpha^{(0)}+\pi^{(0)})\sigma^{(0)} +
  \epsilon^{(0)}\beta^{(0)} -(\bar{\alpha}^{(0)}-\bar{\pi}^{(0)})\epsilon^{(0)} + \Psi_1^{(0)}\,,\\
  D\lambda^{(0)} -\bar{\delta}\pi^{(0)} = -2\epsilon^{(0)}\lambda^{(0)} + 2\alpha^{(0)}\pi^{(0)}\,,\\
  D\mu^{(0)} - \delta\pi^{(0)} = -2\epsilon^{(0)}\mu^{(0)} + 2\beta^{(0)}\pi^{(0)} + \Psi_2^{(0)}\,.
\end{EqSystem*}
Adding Eq.~(\ref{eq:12:b}) to the complex conjugate of
Eq.~(\ref{eq:12:c}), and using the conditions on the spin coefficients
leads to 
\begin{equation}
  \label{eq:18}
  D\alpha^{(0)} = 0\,, \quad D\beta^{(0)} = 0\,,\quad \Psi_1^{(0)} = 0\,.\quad 
\end{equation}
The remaining two equations tell us about the time dependence of the
expansion and shear of $n$ at the horizon:
\begin{EqSystem}[eq:13]
  D\lambda^{(0)} + \tilde{\kappa}\lambda^{(0)} = \bar\delta\pi^{(0)} +
  2\alpha^{(0)}\pi^{(0)}\,,\\ 
  D\mu^{(0)} + \tilde\kappa\mu^{(0)} = \delta\pi^{(0)} + 2\beta^{(0)}\pi^{(0)} + \Psi_2^{(0)}\,.
\end{EqSystem}
Using the $\eth$ operator for the angular derivatives, and noting that
$\pi$ has spin weight $-1$, this is written more cleanly as
\begin{EqSystem*}[eq:21]
  D\lambda^{(0)} + \tilde\kappa\lambda^{(0)} = \bar\eth\pi^{(0)} + (\pi^{(0)})^2\,,\\
  D\mu^{(0)} + \tilde\kappa\mu^{(0)} = \eth\pi^{(0)} + \left|\pi^{(0)}\right|^2 + \Psi_2^{(0)}\,.
\end{EqSystem*}
These two equations can be shown to be identical to Eq.~(3.9) of
\cite{Ashtekar:2001jb} with the matter terms set to zero.  

Now consider the angular field equations (\ref{eq:9}) at the
horizon. The first of these leads to $\Psi^{(0)} = 0$ which we
already knew.  Eq.~(\ref{eq:9:c}) gives
\begin{equation}
  \label{eq:psi3horizon}
  \Psi_3^{(0)} = \bar\eth\mu^{(0)} + \pi^{(0)}\mu^{(0)} - \eth\lambda^{(0)} -\bar{\pi}^{(0)}\lambda^{(0)}\,.
\end{equation}
The real and imaginary parts of Eq.~(\ref{eq:9:b}) give respectively
\begin{eqnarray}
  \label{eq:angulareqs1}
  -2\mathrm{Re}\Psi_2^{(0)} = \delta a + \bar\delta \bar{a}
  -2|a|^2\,,\qquad a:= \alpha^{(0)}-\bar\beta^{(0)}\,,\\
  \label{eq:angulareqs2}
  -2i\mathrm{Im}\Psi_2^{(0)} = \eth\pi^{(0)} - \bar\eth\bar{\pi}^{(0)}\,. 
\end{eqnarray}
Since $a=\alpha^{(0)}-\bar\beta^{(0)}$ determines the connection on the horizon
cross-section, it's derivative must be connected with the scalar
2-curvature of the cross-section.  In fact, it is not hard to show
that the first equation is equivalent to 
\begin{equation} 
  {}^2\tilde{\mathcal{R}} = -4\mathrm{Re}\Psi_2^{(0)}
\end{equation}
The second equation relates the curl of $\tilde{\omega}$ with the
imaginary part of $\Psi_2$ and thus reproduces Eq.~(\ref{eq:16}).

Now the Bianchi identities Eqs.~(\ref{eq:11}) lead to
\begin{EqSystem*}[eq:22]
  D\Psi_2^{(0)}=0\,,\\
  D\Psi_3^{(0)} + \tilde\kappa\Psi_3^{(0)} = \bar\delta\Psi_2^{(0)}
  + 3\pi^{(0)}\Psi_2^{(0)}\,,\\
  D\Psi_4^{(0)} + 2\tilde\kappa\Psi_4^{(0)} = \bar\eth\Psi_3^{(0)} +
  5\pi^{(0)}\Psi_3^{(0)} -3\lambda^{(0)}\Psi_2^{(0)}\,.
\end{EqSystem*}
The second of these can be shown to follow from Eq.~(\ref{eq:psi3horizon}), so
it doesn't yield any new information, but the last equation gives the
time evolution of $\Psi_4$ at the horizon. 

To summarize, here is the free data required at the horizon.  We start
with a choice of null generator $\ell^a$ and a constant surface
gravity $\tilde\kappa := \epsilon^{(0)}+\bar{\epsilon}^{(0)}$, and an affine
parameter $v$.  Surfaces of constant $v$ are spheres $S_v$, and we
choose a basis $m,\bar{m}$ tangent to the $S_v$ and Lie dragged along
$\ell^a$.  Then, on any one of the spheres, say $S_0$, choose the
transversal expansion and shear $\mu^{(0)},\lambda^{(0)}$, the spin
coefficient $\pi^{(0)}$, the connection on $S_0$:
$\alpha^{(0)}-\bar{\beta}^{(0)}$, and the transverse gravitational radiation
$\Psi_4^{(0)}$.  The field equations then show that $\pi^{(0)}$,
$\alpha^{(0)}-\bar{\beta}^{(0)}$ are time independent.  The transversal shear
and expansion $\lambda^{(0)}(v),\mu^{(0)}(v)$ satisfy Eqs.~(\ref{eq:21})
which means that their time evolution is: 
\begin{EqSystem*}
  \mu^{(0)}(v) = \mu^{(0)}(0)e^{-\tilde{\kappa}v} +
  \frac{1}{\tilde\kappa}\left[\bar\eth\pi^{(0)} + (\pi^{(0)})^2\right](1-e^{-\tilde\kappa v})  \,,\\
  \lambda^{(0)}(v) = \lambda^{(0)}(0)e^{-\tilde{\kappa}v} +
  \frac{1}{\tilde\kappa}\left[\eth\pi^{(0)} + \left|\pi^{(0)}\right|^2 + \Psi_2^{(0)}\right](1-e^{-\tilde\kappa v})  \,.
\end{EqSystem*}
The Weyl tensor components $\Psi_0^{(0)},\Psi_1^{(0)}$ vanish.
$\Psi_2^{(0)}$ is time independent, it's real part is determined by
$\alpha^{(0)}-\bar{\beta}^{(0)}$ according to Eq.~(\ref{eq:21}), and
its imaginary part by $\pi^{(0)}=\alpha^{(0)}+\bar{\beta}^{(0)}$
according to Eq.~(\ref{eq:angulareqs2}); $\Psi_3^{(0)}$ is determined
from Eq.~(\ref{eq:psi3horizon}) and its time evolution is of the same
form as for $\lambda^{(0)},\mu^{(0)}$.  From Eq.~(\ref{eq:11:b}) we
get
\begin{equation}
  \Psi_3^{(0)}(v) = \Psi_3^{(0)}(0)e^{-\tilde\kappa v} +
  \frac{1}{\tilde\kappa}\left[ \bar{\eth}\Psi_2^{(0)} + 3\pi^{(0)}\Psi_2^{(0)}\right](1-e^{-\tilde\kappa v})\,.
\end{equation}
Finally, $\Psi_4^{(0)}$ can be freely specified on $S_0$ and it
evolves in time according to Eq.~(\ref{eq:22:c}); this leads to a time
dependence of the form
\begin{equation}
  \Psi_4^{(0)}(v) = \Psi_4^{(0)}(0) +
  \frac{A}{2\tilde\kappa}(1-e^{-2\tilde\kappa v}) +
  \frac{B}{\tilde\kappa}e^{-\tilde\kappa v}(1-e^{-\tilde\kappa v})
\end{equation}
where the time independent angular functions $A$ and $B$ are defined
via 
\begin{eqnarray}
  \left(\bar\eth +
    5\pi^{(0)}\right)\Psi_3^{(0)} -3\lambda^{(0)}\Psi_2^{(0)} = A +
  Be^{-\tilde\kappa v}\,,  \\
  A = \frac{1}{\tilde\kappa}\left[(\bar\eth+ 5\pi^{(0)})(\bar\eth +
    3\pi^{(0)}) - 3\left(\eth\pi^{(0)} +
      \left|\pi^{(0)}\right|^2 +
      \Psi_2^{(0)}\right)\right]\Psi_2^{(0)}\,,\\
  B = (\bar\eth + 5\pi^{(0)})\Psi_3^{(0)}(0) -
  3\lambda^{(0)}\Psi_2^{(0)}  - A\,. 
\end{eqnarray}
We note that the time evolution on an extremal horizon
($\tilde\kappa=0$) would be very different; see
e.g. \cite{Ashtekar:2001jb,Adamo:2009cd}.  As we shall see in the next
section, not only can we specify $\Psi_4$ freely on $S_0$, but
also all of its radial derivatives.  The free data on the transversal
null surface $\mathcal{N}_0$ is in fact $\Psi_4$.

\section{The radial equations}
\label{sec:radial}

With the intrinsic horizon geometry understood, we turn our attention
to the radial equations (\ref{eq:7}) and (\ref{eq:10}) which determine
all radial derivatives for all the non-zero spin coefficients and Weyl
tensor components.  The first radial derivatives are easily obtained
by substituting the horizon values on the right hand sides of
Eqs.~(\ref{eq:7}) and (\ref{eq:10}):
\begin{EqSystem*}[eq:24]
  \kappa^{(1)} = \sigma^{(1)} = 0\,, \\
  \rho^{(1)} = \Psi_2^{(0)}\,,\\
  \epsilon^{(1)} + \bar{\epsilon}^{(1)} = 2\left|\pi^{(0)}\right|^2 +
  2\textrm{Re}[\Psi_2^{(0)}]\,,\\ 
  \epsilon^{(1)}- \bar{\epsilon}^{(1)} = \bar{\pi}^{(0)}(\alpha^{(0)}-\bar{\beta}^{(0)}) -
  \pi^{(0)}(\bar{\alpha}^{(0)}-\beta^{(0)}) + \Psi_2^{(0)} - \bar{\Psi}_2^{(0)}\,,\\ 
  \lambda^{(1)} = 2\mu^{(0)}\lambda^{(0)} + \Psi_4^{(0)}\,,\\
  \mu^{(1)} = (\mu^{(0)})^2 + \left|\lambda^{(0)}\right|^2 \,,\\
  \pi^{(1)} = \alpha^{(1)} + \bar{\beta}^{(1)}  = \pi^{(0)}\mu^{(0)} + \bar{\pi}^{(0)}\lambda^{(0)} + \Psi_3^{(0)}\,,\\
  \alpha^{(1)} -\bar{\beta}^{(1)} = \mu^{(0)}(\alpha^{(0)}-\bar{\beta}^{(0)}) +
  \lambda^{(0)}(\beta^{(0)}-\bar{\alpha}^{(0)}) + \Psi_3^{(0)}\,.\\
  \Psi_0^{(1)} = 0\,,\\
  \Psi_1^{(1)} = -\eth\Psi_2^{(0)}\,,\\
  \Psi_2^{(1)} = -(\eth+\bar{\pi}^{(0)})\Psi_3^{(0)} +
  3\mu^{(0)}\Psi_2^{(0)}\,,\\
  \Psi_3^{(1)} = -(\eth+2\bar{\pi}^{(0)})\Psi_4^{(0)} +
  4\mu^{(0)}\Psi_3^{(0)}\,.
\end{EqSystem*}
We can continue this process by applying the radial derivative
$\Delta$ to Eqs.~(\ref{eq:7}) once again, and substituting the first
derivatives from above on the right hand side. This leads to
expressions for the second derivatives in terms of the horizon data:
\begin{EqSystem*}[eq:26]
  \kappa^{(2)} = -(\eth -\bar{\pi}^{(0)})\Psi_2^{(0)}\,,\\
  \sigma^{(2)} = \bar{\lambda}^{(0)}\Psi_2^{(0)}\,,\\
  \rho^{(2)} = 4\mu^{(0)}\Psi_2^{(0)} - (\eth+\bar{\pi}^{(0)})\Psi_3^{(0)}\,,\\
  \lambda^{(2)} = 6(\mu^{(0)})^2\lambda^{(0)} + 2\lambda^{(0)}\left|\lambda^{(0)}\right|^2 +
  2\mu^{(0)}\Psi_4^{(0)} + \Psi_4^{(1)}\,,\\
  \mu^{(2)} = 2(\mu^{(0)})^3 + 6\mu^{(0)}\left|\lambda^{(0)}\right|^2 +
  \lambda^{(0)}\bar{\Psi}_4^{(0)} + \bar{\lambda}^{(0)}\Psi_4^{(0)}\,,\\
  \pi^{(2)} = 2\pi^{(0)}((\mu^{(0)})^2+\left|\lambda^{(0)}\right|^2) +
  4\bar{\pi}^{(0)}\lambda^{(0)}\mu^{(0)} + 5\mu^{(0)}\Psi_3^{(0)}
  \nonumber \\ \qquad\qquad +
  \lambda^{(0)}\bar{\Psi}_3^{(0)} - (\eth+\bar{\pi}^{(0)})\Psi_4^{(0)}\,,\\
    \alpha^{(2)}-\bar{\beta}^{(2)} = 2(\alpha^{(0)}-\bar{\beta}^{(0)})((\mu^{(0)})^2+
    \left|\lambda^{(0)}\right|^2) +
    4\mu^{(0)}\lambda^{(0)}(\beta^{(0)}-\bar{\alpha}^{(0)}) \nonumber\\
    \qquad\qquad + \Psi_4^{(0)}(\beta^{(0)}-\bar{\alpha}^{(0)}) + 5\mu^{(0)}\Psi_3^{(0)} -
    \lambda^{(0)}\bar{\Psi}_3^{(0)} - (\eth+2\bar{\pi}^{(0)})\Psi_4^{(0)}\,.
\end{EqSystem*}
Finally we can also investigate the non-radial equations to
$\mathcal{O}(r)$.  For this, we need the directional derivatives to
$\mathcal{O}(r)$ which we have already calculated:
\begin{EqSystem}
  \Delta = -\frac{\partial}{\partial r}\,,\\
  D = \frac{\partial}{\partial v} + r(\epsilon^{(0)} +
  \bar{\epsilon}^{(0)})\frac{\partial}{\partial r} +
  r\pi^{(0)}\xi^i_{(0)}\frac{\partial}{\partial x^i} +
  r\bar{\pi}^{(0)}\bar{\xi}^i_{(0)}\frac{\partial}{\partial x^i} + \mathcal{O}(r^2)\,,\\
  \delta = \xi^i_{(0)}\frac{\partial}{\partial x^i} +
  r\mu^{(0)}\xi^i_{(0)}\frac{\partial}{\partial x^i}+
  r\bar{\lambda}^{(0)}\bar{\xi}^i_{(0)}\frac{\partial}{\partial x^i} +
  r\bar{\pi}^{(0)}\frac{\partial}{\partial r}  + \mathcal{O}(r^2)\,.
\end{EqSystem}
We substitute these on the left hand sides of
Eqs.~(\ref{eq:frametime}), (\ref{eq:8}) and (\ref{eq:9}), and also
expand the right hand sides in powers of $r$.  Equating powers of $r$
on both sides yields differential equations at each order.  The
$\mathcal{O}(r^0)$ terms have already been studied in
Sec.~\ref{sec:horizon} since these deal with the intrinsic horizon
geometry.  The $\mathcal{O}(r)$ equations are straightforward in some
cases.  For example, Eq.~(\ref{eq:dsigma}) is trivial because
$\kappa,\sigma,\Psi_0$ are all $\mathcal{O}(r^2)$ quantities as we have
already seen in Eqs.~(\ref{eq:24}).  Similarly, Eq.~(\ref{eq:drho}) is
also easy. Using $\rho = r\Psi_2^{(0)} + \mathcal{O}(r^2)$ and the
vanishing of $\kappa,\sigma$ to $\mathcal{O}(r)$, we get 
\begin{equation}
  \label{eq:1}
  \frac{\partial}{\partial v}\Psi_2^{(0)} = 0\,,
\end{equation}
which we already knew.  The remaining equations are similarly
straightforward, but a lot more tedious.  In the end, it turns out
that Eqs.~(\ref{eq:frametime}), (\ref{eq:8}) and (\ref{eq:9}) do not
yield any new information at $\mathcal{O}(r)$.

\section{The expansion of the metric}
\label{sec:metric}

To expand the metric in powers of $r$, we start with the frame fields
defined in Eqs.(\ref{eq:2}) and (\ref{eq:3}), and the radial frame
equations (\ref{eq:4}) derived from the commutation relations.  The
strategy is the same as for the spin coefficients.  Eqs.~(\ref{eq:4})
give us the first radial derivatives by substituting the horizon
values on the right hand side, and taking higher derivatives leads to
the higher order terms. The calculations are straightforward and lead
to the following expansions: 
\begin{EqSystem*}[eq:27]
  U = r\tilde{\kappa} + r^2\left(2\left|\pi^{(0)}\right|^2 +
    \mathrm{Re}[\Psi_2^{(0)}]\right) + \mathcal{O}(r^3)\,, \\
  \Omega = r\bar{\pi}^{(0)} + r^2\left( \mu^{(0)}\bar{\pi}^{(0)} +
    \bar{\lambda}^{(0)}\pi^{(0)} + \frac{1}{2}\bar{\Psi}_3^{(0)}\right) +
  \mathcal{O}(r^3)\,, \\
  X^i = \bar{\Omega} \xi^i_{(0)}  + \Omega\bar{\xi}^i_{(0)} + \mathcal{O}(r^3)\,,\\
  \xi^i = \left[ 1 + r\mu^{(0)} + r^2\left((\mu^{(0)})^2+|\lambda^{(0)}|^2\right)\right]\xi^i_{(0)}
  \nonumber \\ \qquad + \left[ r\bar{\lambda}^{(0)}  + r^2\left( 2\mu^{(0)}\bar{\lambda}^{(0)} +
      \frac{1}{2}\bar{\Psi}_4^{(0)}\right) \right]\bar{\xi}^i_{(0)}  +
  \mathcal{O}(r^3)\,.
\end{EqSystem*}
The contravariant metric is seen to be given in terms of the frame fields
as follows:
\begin{EqSystem*}[eq:28]
  g^{rr} = 2(U+|\Omega|^2) \,,\\
  g^{vr} = 1\,,\\
  g^{ri} = X^i + \bar{\Omega}\xi^i + \Omega\bar{\xi}^i \\
  g^{ij} = \xi^i\bar{\xi}^j + \bar{\xi}^i\xi^j\,. 
\end{EqSystem*}
An explicit calculation yields
\begin{MultiLine}
  g^{rr} = 2\tilde{\kappa}r + 2r^2\left(3|\pi^{(0)}|^2 +
    \mathrm{Re}[\Psi_2^{(0)}]\right) + \mathcal{O}(r^3) \,.
\end{MultiLine}
\begin{MultiLine}
  g^{ir} = \left[ 2r\pi^{(0)} + r^2\left(2\mu^{(0)}\pi^{(0)} +
      2\lambda^{(0)}\bar{\pi}^{(0)} + \frac{1}{2}\Psi_3^{(0)}\right) + \mathcal{O}(r^3)\right]\xi^i_{(0)} \nonumber\\*
  + \left[2r\bar{\pi}^{(0)} + r^2\left( 2\mu^{(0)}\bar{\pi}^{(0)} +
      2\bar{\lambda}^{(0)}\pi^{(0)} + \frac{1}{2}\bar{\Psi}_3^{(0)} \right) +
    \mathcal{O}(r^3) \right]\bar{\xi}^i_{(0)}\,.
\end{MultiLine}
\begin{MultiLine}
  g^{ij} = \left(1+2r\mu^{(0)} +
    3r^2((\mu^{(0)})^2+|\lambda^{(0)}|^2)\right)(\xi^i_{(0)}\bar{\xi}^j_{(0)} +
  \bar{\xi}^i_{(0)}\xi^j_{(0)}) \\* 
  + \left[r\lambda^{(0)} + r^2 \left(3\mu^{(0)}\lambda^{(0)}+
      \frac{1}{2}\Psi_4^{(0)}\right)\right]\,2\xi^i_{(0)}\xi^j_{(0)} \\*
 + \left[r\bar{\lambda}^{(0)} + r^2 \left(3\mu^{(0)}\bar{\lambda}^{(0)}+
     \frac{1}{2}\bar{\Psi}_4^{(0)}\right)\right]\,2\bar{\xi}^i_{(0)}\bar{\xi}^j_{(0)}
 +   \mathcal{O}(r^3)\,.
\end{MultiLine}
The null co-tetrad, i.e. the dual basis for the null tetrad found
above can also be calculated easily up to $\mathcal{O}(r^2)$ :
\begin{EqSystem}
  n = -dv\,,\\
  \ell = dr - \left(\tilde{\kappa}r +
    \mathrm{Re}[\Psi_2^{(0)}]r^2\right)dv - \left(\pi^{(0)} r +
    \frac{1}{2}\Psi_3^{(0)}r^2\right)\xi^{(0)}_idx^i - \left(\bar{\pi}^{(0)} r
    +
    \frac{1}{2}\bar{\Psi}_3^{(0)}r^2\right)\bar{\xi}^{(0)}_idx^i\,,\\
  m = -\left(\bar{\pi}^{(0)}r+ \frac{1}{2}\Psi_3^{(0)}r^2\right)dv +
  (1-\mu^{(0)}r)\xi^{(0)}_idx^i - \left(\bar{\lambda}^{(0)}r +
    \frac{1}{2}\bar{\Psi}_4^{(0)}r^2\right)\bar{\xi}^{(0)}_idx^i\,. 
\end{EqSystem}
Here $\xi^{(0)}_i$ are defined by the relations $\xi^{(0)}_i\xi^i_{(0)} = 0$ and
$\xi^{(0)}_i\bar{\xi}^i_{(0)} = 1$; it will be convenient to set $m^{(0)}_a :=
\xi^{(0)}_i\partial_ax^i$.  In powers of $r$, the metric is:
\begin{equation}
  \label{eq:metric-expansion}
  g_{ab} = -2\ell_{(a}n_{b)} + 2m_{(a}\bar{m}_{b)} = g_{ab}^{(0)} +
  g_{ab}^{(1)}r + \frac{1}{2}g_{ab}^{(2)}r^2 + \cdots\,,
\end{equation}
where
\begin{Equation}
  \label{eq:metric0}
  g_{ab}^{(0)} = 2\partial_{(a}r\partial_{b)}v +
  2m^{(0)}_{(a}\bar{m}^{(0)}_{b)}\,,
\end{Equation}
\begin{MultiLine}
  \label{eq:metric1}
  g_{ab}^{(1)} =  -\left( 2\tilde{\kappa}\partial_{(a}v\partial_{b)}v +
    4\pi^{(0)}m^{(0)}_{(a}\partial_{b)}v +
    4\bar{\pi}^{(0)}\bar{m}^{(0)}_{(a}\partial_{b)}v \right. \\ +
    \left. 4\mu^{(0)}m^{(0)}_{(a}\bar{m}^{(0)}_{b)} +
    2\lambda^{(0)}m^{(0)}_{(a}m^{(0)}_{b)} +
    2\bar{\lambda}^{(0)}\bar{m}^{(0)}_{(a}\bar{m}^{(0)}_{b)}\right)\,,
\end{MultiLine}
\begin{MultiLine}
  \label{eq:metric2}
  g_{ab}^{(2)} = 4\left(|\pi^{(0)}|^2 -
    \mathrm{Re}[\Psi_2^{(0)}]\right)\partial_{(a}v\partial_{b)}v \\*
  + 4\left( \mu^{(0)}\pi^{(0)} + \lambda^{(0)}\bar{\pi}^{(0)}
    -\Psi_3^{(0)}\right)m^{(0)}_{(a}\partial_{b)}v \\*
  +4\left( \mu^{(0)}\bar{\pi}^{(0)} + \bar{\lambda}^{(0)}\pi^{(0)} -
    \bar{\Psi}_3^{(0)}\right)\bar{m}^{(0)}_{(a}\partial^{}_{b)}v \\*
  + 4\left((\mu^{(0)})^2+\left|\lambda^{(0)}\right|^2 \right)
  m^{(0)}_{(a}\bar{m}^{(0)}_{b)} \\*
  + \left(4\mu^{(0)}\lambda^{(0)} - 2\Psi_4^{(0)}
  \right)m^{(0)}_{(a}m^{(0)}_{b)} +
  \left(4\mu^{(0)}\bar{\lambda}^{(0)} -2\bar{\Psi}_4^{(0)}
  \right)\bar{m}^{(0)}_{(a}\bar{m}^{(0)}_{b)}\,.
\end{MultiLine}
We could easily obtain the third order metric by using
Eqs.~(\ref{eq:26}), but we shall not do so here.

\section{Inclusion of electromagnetic fields}
\label{sec:maxwell}

Let us now go beyond vacuum spacetimes and include electromagnetic
fields.  As usual, we expand the Maxwell field components $\phi_k$ in
powers of $r$:
\begin{equation}
  \phi_k = \phi^{(0)}_k + r\phi^{(1)}_k + \frac{1}{2}\phi^{(2)}_k +
  \ldots\,,\quad \textrm{where}\quad k=0,1,2\,.
\end{equation}
Consider first the intrinsic horizon geometry.  We start with the
behavior of the Maxwell field components at the horizon. The
Raychaudhuri equation for the null generators, after imposing the
conditions of Eqs.~(\ref{eq:13}) and (\ref{eq:6}) is just one of the
Newman-Penrose field equation (Eq.~\ref{eq:drho}) that we have used
earlier.  In the presence of a Maxwell field, that equation becomes
\begin{equation} 
  D\rho -\bar{\delta}\kappa = \rho^2 + |\sigma|^2 +
  (\epsilon+\bar{\epsilon})\rho -2\alpha\kappa + 2|\phi_0|^2\,.\label{eq:drhoem}\\ 
\end{equation}
On the horizon, by definition $\rho = \kappa =0$.  Thus from the above
equation we deduce that we still have $\sigma=0$ as before, and in addition we
also get 
\begin{equation}
  \phi^{(0)}_0 = 0\,.
\end{equation}
This implies that 
\begin{equation}
  \Phi^{(0)}_{00} = \Phi^{(0)}_{01} = \Phi^{(0)}_{02} =
  \Phi^{(0)}_{10} = \Phi^{(0)}_{20} = 0\,. 
\end{equation}
Furthermore, we also have $\Lambda=0$ everywhere since the trace of
$T_{ab}$ vanishes.  Imposing these conditions to determine the time
evolution of the geometrical fields on the horizon, the two non-radial
Maxwell equations (\ref{eq:maxwelleqs:a}) and (\ref{eq:maxwelleqs:b})
become
\begin{EqSystem*}[eq:maxwelltime]
  D\phi_1^{(0)} = 0\,,\\
  D\phi_2^{(0)} + \tilde{\kappa}\phi_2^{(0)} =
  \bar{\delta}\phi_1^{(0)} + 2\pi^{(0)}\phi_1^{(0)} \,.
\end{EqSystem*}
Thus, $\phi_1^{(0)}$ is time independent, the charges defined in
Eq.~(\ref{eq:emcharges}) are constant. Since $\phi_0^{(0)}$ vanishes,
these charges are in fact independent of which cross-section of
$\Delta$ is used in the integral. Just like $\Psi_3$ and $\Psi_4$ at
the horizon, $\phi_2$ has a time dependence.

With these simplifications at hand, it turns out that all of
Eqs.~(\ref{eq:12}) remain unchanged.  The only change in the intrinsic
geometry occurs in two of the angular equations (\ref{eq:9}).  The
first change is an extra contribution in Eq.~(\ref{eq:angulareqs1}):
\begin{equation}
  -2\mathrm{Re}[\Psi_2^{(0)}] + 4|\phi_1^{(0)}|^2 = \delta a + \bar\delta \bar{a}
  -2|a|^2 \Rightarrow  -4\mathrm{Re}[\Psi_2^{(0)}] + 8|\phi_1^{(0)}|^2 = {}^2\tilde{\mathcal{R}}\,.
\end{equation}
The second effect is in Eq.~(\ref{eq:psi3horizon}) which determines
$\Psi_3^{(0)}$:  
\begin{equation}
  \label{eq:psi3horizonem}
  \Psi_3^{(0)} = \bar\eth\mu^{(0)} + \pi^{(0)}\mu^{(0)} - \eth\lambda^{(0)} -
  \bar{\pi}^{(0)}\lambda^{(0)} + 2\phi_2^{(0)}\bar{\phi}_1^{(0)}\,.
\end{equation}

The remaining Maxwell equations (\ref{eq:maxwelleqs:c}) and
(\ref{eq:maxwelleqs:d}) determine the radial derivatives of
$\phi_0$ and $\phi_1$.  As usual, we impose Eqs.~(\ref{eq:13}) and
(\ref{eq:6}) to obtain the radial equations for $\phi_0$ and $\phi_1$:
\begin{EqSystem*}[eq:maxwellradial]
  \Delta\phi_0 = \delta\phi_1 - \mu\phi_0 + \sigma\phi_2\,,\\
  \Delta\phi_1 = \delta\phi_2 - 2\mu\phi_1 + 2\beta\phi_2\,.
\end{EqSystem*}
Note that there is no equation for the radial derivative of $\phi_2$.
Thus, just like $\Psi_4$, we need to specify $\psi_2$ on the past
light cone $\mathcal{N}_0$ as part of the free data.  As for the other
radial equations, we can iterate this to obtain an expansion for
$\phi_0$ and $\phi_1$.  To first order we need to simply evaluate the
right hand side of these equations at the horizon: 
\begin{EqSystem*}
  \phi_0^{(1)} = -\delta\phi_1^{(0)}\,,\\
  \phi_1^{(1)} = -\delta\phi_2^{(0)} + 2\mu^{(0)}\phi_1^{(0)} +
  2\beta^{(0)}\phi_2^{(0)} = (-\eth + \pi^{(0)})\phi_2^{(0)} + 2\mu^{(0)}\phi_1^{(0)}\,.
\end{EqSystem*}
Let us turn now to the Bianchi identities.  In the vacuum case, we had
the Equations (\ref{eq:10}) and (\ref{eq:11}).  These equations give
here as well the radial and time derivatives of the Weyl tensor
components with additional terms for the matter contributions.  For
the time derivatives at the horizon, we only need to worry about
$\Psi_4^{(0)}$ (since $\Psi_3^{(0)}$ has been determined above and the
others are time independent): 
\begin{MultiLine}
  D\Psi_4^{(0)} + 2\tilde\kappa\Psi_4^{(0)} &= (\bar\eth +
  5\pi^{(0)})\Psi_3^{(0)} + 2\bar\phi_1^{(0)} (\bar\eth +
  \pi^{(0)})\phi_2^{(0)} \\ - 3\lambda^{(0)}\Psi_2^{(0)} -
  4\lambda^{(0)}|\phi_1^{(0)}|^2\,.  
\end{MultiLine}
The radial equations for the Weyl tensor components from the Bianchi
identities are not very enlightening, and we shall just write the
first derivatives evaluated at the horizon: 
\begin{EqSystem*}
  \Psi_0^{(1)} = 0\,,\\
  \Psi_1^{(1)} = -\delta\Psi_2^{(0)} - 2\bar\phi_1^{(0)}\delta\phi_1^{(0)} \,,\\
  \Psi_2^{(1)} = -(\eth + \bar\pi^{(0)})\Psi_3^{(0)} -
  2\bar\phi_1^{(0)}(\eth+\bar\pi^{(0)})\phi_2^{(0)} + 2\bar\phi_2^{(0)}\bar\delta\phi_1^{(0)} +
  3\mu\Psi_2^{(0)}  + 4\mu|\phi_1^{(0)}|^2 \,,\\
  \Psi_3^{(1)} = -(\eth + 2\pi^{(0)})\Psi_4^{(0)} +
  4\mu^{(0)}\Psi_3^{(0)} + 2\bar\phi_1^{(0)}\phi_2^{(1)} +
  2\bar\phi_2^{(0)}(\eth - 2\pi^{(0)})\phi_2^{(0)}\nonumber \\\qquad 
  -4\lambda^{(0)}\phi_1^{(0)}\bar\phi_2^{(0)} + 4\pi^{(0)}|\phi_2^{(0)}|^2
\end{EqSystem*}
In the presence of matter fields, there are three (two real and one
complex) additional Bianchi identities containing only Ricci tensor
components.  Let us consider these equations at the horizon after
imposing our conditions on the spin coefficients and the vanishing of
$\phi_0$.  The first of these equations yields simply that
$\phi_0^{(0)}$ is time independent (which we already knew):
\begin{equation}
  D\Phi_{11}^{(0)} = 0 \Rightarrow D\phi_1^{(0)} = 0\,.
\end{equation}
The second determines the time evolution of $\Phi_{12}^{(0)}$ 
\begin{equation}
  D\Phi_{12}^{(0)} - \delta\Phi_{11}^{(0)} + \Delta\Phi_{01}^{(0)} =
  2\bar\pi^{(0)} \Phi_{11}^{(0)} - 2\bar{\epsilon}^{(0)}\Phi_{12}^{(0)} \,.
\end{equation}
The third yields the time dependence of $\Phi_{22}$:
\begin{MultiLine}
  D\Phi_{22}^{(0)} - \delta\Phi_{21}^{(0)} + \Delta\Phi_{11}^{(0)} -
  \bar{\delta}\Phi_{12}^{(0)} & = -4\mu^{(0)}\Phi_{11}^{(0)} + (2\pi^{(0)} +
  2\bar{\beta}^{(0)})\Phi_{12}^{(0)} \\ + (2\bar\pi^{(0)} +
  2\beta^{(0)})\Phi_{21}^{(0)} - 4\epsilon^{(0)}\Phi_{22}^{(0)}\,.    
\end{MultiLine}
Using $\Phi_{ij} = 2\phi_i\bar{\phi}_j$, it can be shown that these
two equations do not have any new information and can be obtained by
combining Eqs.~(\ref{eq:maxwelltime}) and (\ref{eq:maxwellradial}).  

In the rest of this section, let us us focus on the expansion of the
metric up to $\mathcal{O}(r^2)$.  First we need the radial derivatives
of the spin coefficients as in Eqs.~(\ref{eq:24}). Let us rewrite
Eqs.~(\ref{eq:24}) keeping this time the contributions of $\phi_1$ and
$\phi_2$ at the horizon, and focusing on the spin coefficients which
is all we need for expanding the metric up to second order:
\begin{EqSystem*}[eq:29]
  \kappa^{(1)} = \sigma^{(1)} = 0\,, \\
  \rho^{(1)} = \Psi_2^{(0)}\,,\\
  \epsilon^{(1)} + \bar{\epsilon}^{(1)} = 2\left|\pi^{(0)}\right|^2 +
  2\textrm{Re}[\Psi_2^{(0)}] + 4|\phi_1^{(0)}|^2\,,\\
  \epsilon^{(1)}- \bar{\epsilon}^{(1)} = \bar{\pi}^{(0)}(\alpha^{(0)}-\bar{\beta}^{(0)}) -
  \pi^{(0)}(\bar{\alpha}^{(0)}-\beta^{(0)}) + \Psi_2^{(0)} - \bar{\Psi}_2^{(0)}\,,\\
  \lambda^{(1)} = 2\mu^{(0)}\lambda^{(0)} + \Psi_4^{(0)}\,,\\
  \mu^{(1)} = (\mu^{(0)})^2 + \left|\lambda^{(0)}\right|^2  + 2\|\phi_2^{(0)}|^2\,,\\
  \pi^{(1)} = \alpha^{(1)} + \bar{\beta}^{(1)} = \pi^{(0)}\mu^{(0)} +
  \bar{\pi}^{(0)}\lambda^{(0)} + \Psi_3^{(0)} + 2\phi_2^{(0)}\bar{\phi}_1^{(0)}\,,\\
  \alpha^{(1)} -\bar{\beta}^{(1)} = \mu^{(0)}(\alpha^{(0)}-\bar{\beta}^{(0)}) +
  \lambda^{(0)}(\beta^{(0)}-\bar{\alpha}^{(0)}) + \Psi_3^{(0)}+ 2\phi_2^{(0)}\bar{\phi}_1^{(0)}\,.
\end{EqSystem*}
We now need the radial derivatives for the frame functions given in
Eqs.~(\ref{eq:4}) which leads us to the expansion of the frame functions:
\begin{EqSystem*}[eq:30]
  U = r\tilde{\kappa} + r^2\left(2\left|\pi^{(0)}\right|^2 +
    \mathrm{Re}[\Psi_2^{(0)}] + 2|\phi_1^{(0)}|^2\right) + \mathcal{O}(r^3)\,, \\
  \Omega = r\bar{\pi}^{(0)} + r^2\left( \mu^{(0)}\bar{\pi}^{(0)} +
    \bar{\lambda}^{(0)}\pi^{(0)} + \frac{1}{2}\bar{\Psi}_3^{(0)} +
    \phi_1^{(0)}\bar{\phi}_2^{(0)}\right) +
  \mathcal{O}(r^3)\,, \\
  X^i = \bar{\Omega} \xi^i_{(0)}  + \Omega\bar{\xi}^i_{(0)} + \mathcal{O}(r^3)\,,\\
  \xi^i = \left[ 1 + r\mu^{(0)} +
    r^2\left((\mu^{(0)})^2+|\lambda^{(0)}|^2 +
      |\phi_2^{(0)}|^2\right)\right]\xi^i_{(0)} \nonumber \\ \qquad +
  \left[ r\bar{\lambda}^{(0)} + r^2\left(
      2\mu^{(0)}\bar{\lambda}^{(0)} +
      \frac{1}{2}\bar{\Psi}_4^{(0)}\right) \right]\bar{\xi}^i_{(0)} + \mathcal{O}(r^3)\,.
\end{EqSystem*}
It is instructive to compare this with Eqs.~(\ref{eq:27}).  We see
that the electromagnetic field modifies $\Psi_2$ and $\Psi_3$ according
to 
\begin{equation}
  \mathrm{Re}[\Psi_2^{(0)}] \rightarrow \mathrm{Re}[\Psi_2^{(0)}] +
  2|\phi_1^{(0)}|^2\,,\qquad \Psi_3^{(0)} \rightarrow \Psi_3^{(0)} +
  2\phi_1^{(0)}\bar{\phi}_2^{(0)}\,. 
\end{equation}
In addition, the $(\mu^{(0)})^2 + |\lambda^{(0)}|^2$ term gets an additive
correction by $|\phi_2^{(0)}|^2$.  

We can finally write the expansion of the metric.  It is clear that
there is no change up to $\mathcal{O}(r)$.  Just like the Weyl tensor,
the effects of the electromagnetic field first enter the metric at
second order.  Thus, Eqs.~(\ref{eq:metric0}) and (\ref{eq:metric0})
are unchanged and Eq.~(\ref{eq:metric2}) becomes:
\begin{MultiLine}
  g_{ab}^{(2)} &= 4\left(|\pi^{(0)}|^2 - \mathrm{Re}[\Psi_2^{(0)}] +
    2|\phi_1^{(0)}|^2\right)\partial_{(a}v\partial_{b)}v \\
  + 4\left( \mu^{(0)}\pi^{(0)} + \lambda^{(0)}\bar{\pi}^{(0)}
    -\Psi_3^{(0)} - 2\phi_2^{(0)}\bar{\phi}_1^{(0)}\right)m^{(0)}_{(a}\partial_{b)}v \\
  +4\left( \mu^{(0)}\bar{\pi}^{(0)} + \bar{\lambda}^{(0)}\pi^{(0)} -
    \bar{\Psi}_3^{(0)} -
    2\bar{\phi}_2^{(0)}\phi_1^{(0)}\right)\bar{m}^{(0)}_{(a}\partial^{}_{b)}v
  \\ + 4\left((\mu^{(0)})^2+\left|\lambda^{(0)}\right|^2 + |\phi_2^{(0)}|^2\right)
  m^{(0)}_{(a}\bar{m}^{(0)}_{b)} \\
  + \left(4\mu^{(0)}\lambda^{(0)} - 2\Psi_4^{(0)}
  \right)m^{(0)}_{(a}m^{(0)}_{b)} + \left(4\mu^{(0)}\bar{\lambda}^{(0)}
    -2\bar{\Psi}_4^{(0)} \right)\bar{m}^{(0)}_{(a}\bar{m}^{(0)}_{b)}\,.
\end{MultiLine}

\section{Conclusions}
\label{sec:conc}

In this paper we have investigated the geometry in the vicinity of a
non-extremal weakly isolated horizon using the Newman-Penrose
formalism. We allow the horizon to have generic values of mass, spin
and charge and higher multipole moments.  We have obtained the
connection, metric and the Weyl tensor up to $\mathcal{O}(r^2)$.  On
the horizon, we choose a parametrization of the null generators so
that we get a specific value of the surface gravity.  In addition, on
an initial spherical cross section $S_0$ of the horizon, we specify
the spin coefficients $\pi, \alpha-\bar{\beta}, \mu, \lambda$.  The
spin coefficients $\pi$ and $\alpha-\bar{\beta}$ out to be time
independent and they specify respectively the $\mathrm{Im}[\Psi_2]$
and the scalar curvature of $S_0$, and thus the spin and mass
multipole moments of the horizon respectively.  The transversal
expansion and shear $\mu,\lambda$ turn out to be time dependent and
they determine $\Psi_3$. In addition, we need to specify $\Psi_4$ on
the past-outgoing light cone $\mathcal{N}_0$ originating from $S_0$.
If we consider charged horizons, then we must specify in addition the
Maxwell field component $\phi_1$ on $S_0$ (this determines the
electric and magnetic charges of the horizon) and $\phi_2$ on
$\mathcal{N}_0$.

We emphasize once again that this is a local calculation and not all
solutions produced by this construction are physically meaningful; the
interesting issue of global existence of solutions has not been
addressed here. For example, the uniqueness theorems would imply that
not all values of the horizon multipole moments would lead to
asymptotically flat solutions.  However, the set of solution we have
obtained encompasses all physically interesting situations (in so far
as the assumption of the horizon being isolated is valid).

There are a number of possible extensions of this work.  The first is
the study of tidally deformed black holes which has, among other
things, implications for the the equation of motion and self force in
general relativity \cite{Poisson:2003nc}.  This requires us to
consider Kerr black holes perturbed by a background curvature, and to
interpret the resulting metric in terms of a moving point particle.
This will be studied in a forthcoming paper.  The near horizon
geometry computed here could be matched to an appropriate far zone
metric to obtain initial data for numerical simulations of binary
black hole systems (see e.g. \cite{Yunes:2006iw}) and finally, future
gravitational wave observatories might be able to measure the tidal
deformations through observations of signals from extreme mass ratio
systems. Knowledge of the near horizon geometry including deviations
from the Kerr multipoles values (which are allowed here) will be
useful in the searches for these signals and in testing the Kerr
nature of the black hole.  Finally beyond the specific calculation of
the near horizon metric, the characteristic initial value
formulation based on the horizon might be useful for wave extraction
in numerical relativity signals.  If $S_0$ is taken to be a
cross-section of the horizon at very late times, then the light cone
$\mathcal{N}_0$ can be viewed as an approximate null infinity.  The
usefulness of this construction remains to be explored.

\section*{Acknowledgments}

I thank Abhay Ashtekar for valuable suggestions and discussions.  I
am also grateful to Bernard Schutz, Toshifumi Futamase, Dirk
P\"utzfeld and Stas Babak for discussions regarding possible
applications of this work.  

\section*{References}

\bibliography{ihbib}{}
\bibliographystyle{vancouver}

\end{document}